\documentclass[runningheads]{llncs}
\usepackage[T1]{fontenc}
\usepackage{subfigure}
\usepackage{graphicx}
\usepackage{multirow}
\usepackage{CJKutf8}
\begin{document}
\title{Achieving Timestamp Prediction While Recognizing with Non-Autoregressive End-to-End ASR Model}
\titlerunning{Timestamp Prediction with NAR Model}
%
\author{Xian Shi \and
Yanni Chen \and
Shiliang Zhang \and
Zhijie Yan}
\authorrunning{X. Shi et al.}
%
\institute{Speech Lab, Alibaba Group, China\\
\email{\{shixian.shi, cyn244124, sly.zsl, zhijie.yzj\}@alibaba-inc.com}}
%
\maketitle              
\begin{abstract}
Conventional ASR systems use frame-level phoneme posterior to conduct force-alignment~(FA) and  provide timestamps, while end-to-end ASR systems especially AED based ones are short of such ability. This paper proposes to perform timestamp prediction~(TP) while recognizing by utilizing continuous integrate-and-fire~(CIF) mechanism in non-autoregressive ASR model - Paraformer. Foucing on the fire place bias issue of CIF, we conduct post-processing strategies including fire-delay and silence insertion. Besides, we propose to use scaled-CIF to smooth the weights of CIF output, which is proved beneficial for both ASR and TP task. Accumulated averaging shift~(AAS) and diarization error rate~(DER) are adopted to measure the quality of timestamps and we compare these metrics of proposed system and conventional hybrid force-alignment system. The experiment results over manually-marked timestamps testset show that the proposed optimization methods significantly improve the accuracy of CIF timestamps, reducing 66.7\% and 82.1\% of AAS and DER respectively. Comparing to Kaldi force-alignment trained with the same data, optimized CIF timestamps achieved 12.3\% relative AAS reduction.

\end{abstract}
\section{Introduction}

Timestamp prediction is one of the most important and widely used subtasks of automatic speech recognition~(ASR). Kinds of speech related tasks~(text-to-speech, key-word spotting)~\cite{t1,t2}, speech/language analysis~\cite{a1,a2,a3} and ASR training strategies~\cite{s1} can be conducted with a reliable timestamp predicting system. At present, conventional hybrid HMM-GMM or HMM-DNN systems are mostly used to conduct force-alignment~(FA) - the given transcriptions are expanded to phoneme sequences for processing Viterbi decoding in WFST~(weighted finite state transducers) composed by acoustic model, lexicon and language model. Recent years have seen the rapid growth of end-to-end~(E2E) ASR models~\cite{ctc,rnnt,tsfm,las}, which skip the complex training process and preparation of language related expert knowledge and convert speech to text with a single neural network. Generally, E2E ASR models are classified into two categorise: Time-synchronous models including CTC and RNN-Transducer, token-synchronous models including listen-attend-and-spell~(LAS) and Transformer based AED models. The former utilize CTC-like criterion over frame-level encoder output and predict posterior probabilities of tokens together with blank while the later conduct cross-attention between acoustic information and character information to achieve soft alignment. These models have shown strong competitiveness and have replaced conventional ASR model in many scenarios. At the same time, however, due to the inherent deficiency of timestamp prediction ability of these models, some ASR systems have to use an additional conventional ASR model to predict the timestamp of recognition results, which introduce computation overhead and training difficulty.

In this paper, we propose to achieve timestamp prediction while recognizing with non-autoregressive E2E ASR model Paraformer~\cite{paraformer}. Continuous integrate-and-fire (CIF)~\cite{cif} is a soft and monotonic alignment mechanism proposed for E2E ASR which is adopted by Paraformer as predictor, it predicts the number of output tokens by integrating frame-level weights, once the accumulated weights exceed the fire threshold, the encoder output of these frames will be summed up to one step of acoustic embedding. One of the core ideas of achieving non-autoregressive decoding in Paraformer is to generate character embedding which has the same length as output sequence. The modeling characteristic of Parafirner CIF delights us to conduct timestamp prediction basing on CIF output. Focusing on the distribution of original CIF inside Paraformer, we propose scaled-CIF training strategies and three post-processing methods to achieve timestamp prediction of high quality and also explore to measure the timestamp prediction systems with AAS and DER metrics. The following part of this paper is organized as below. Timestamp prediction and FA related works are introduced in section 2. We briefly introduce CIF and Paraformer and look into original CIF distribution in section 3. Section 4 comes about the proposed methods including scaled-CIF and post-processing strategies while section 5 describe out experiments and results in detail. Section 6 ends the paper with our conclusion.

\vspace{2mm}
\noindent\textbf{Our contributions are:}
\begin{itemize}
    \item From the aspects of timestamp quality, the proposed scaled-CIF and post-processing strategies improve the accuracy of timestamp and outperforms the conventional hybrid model trained with the same data.
    \item This paper propose to predict timestamps naturally while recognizing with Paraformer, such system can predict accurate timestamp of recognition results and reduce computation overhead which is of value in commercial usage.
\end{itemize}

\section{Related Works}

\begin{figure}[htbp]
    
    \includegraphics[width=\linewidth]{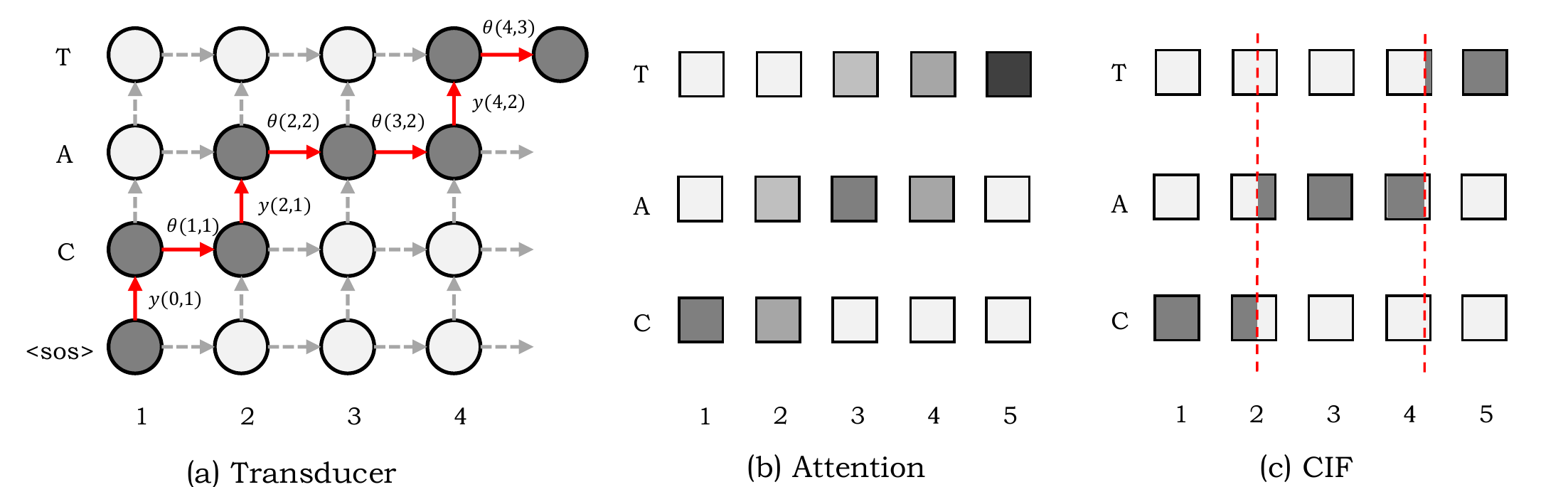}
    \caption{Illustration of alignment of transducer, attention and CIF.}\label{align}
\end{figure}

In this section we briefly introduce the mechanism of sequence-to-sequence modeling and discuss about the recent works related to timestamp predicton and force-alignment. Time-synchronous models and token-synchronous solve unequal length sequence prediction in different ways. Transducer~\cite{rnnt} performs forward and backward algorithm as shown in~\ref{align} (a), it is allowed to move in time axis or label axis to establish connection between token sequence and time sequence of unequal length. However, it turns out that a well trained CTC~\cite{ctc} or Transducer model tends to predict posterior probabilities with sharp peaks~(single frame with extremely high probability for a token except blank)~\cite{ctcpeak}, and the position of the peak can not reflect the real time of the token, especially when the modeling units have long duration. AED based E2E models like Transformer conduct cross-attention between encoder and decoder, the score matrix inside cross-attention can be regarded as alignment but it is soft and nonmonotonic, so it's hard to conduct timestamp prediction in origin AED-based models like Transformer. CIF is a time-synchronous method which is also adopted by AED based models~\cite{paraformer,bdcif}. It generates monotonic alignment by predicting integrate weights in frame level which can be naturally treated as timestamps.

Recently, neural network based timestamp prediction and force-alignment strategies without conventional hybrid systems are explored. Kürzinger et al.~\cite{ctcseg} proposed CTC-segmentation for German speech recognition, which determines the alignment through forward and backward probabilities of CTC model. The proposed alignment system beats the conventional ones including HTK~\cite{htk} and Kaldi~\cite{kaldi} in German task. Li et al.~\cite{neufa} proposed NEUFA to conduct force-alignment, which deploys bidirectional attention mechanism to achieve bidirectional relation learning for parallel text and speech data. The proposed boundary detector
takes the attention weights from both ASR and TTS directions as inputs to predict the left and right boundary signals for each phoneme. Their system achieves better accuracies at different tolerances comparing to MFA~\cite{mfa}. Besides, systems like ITSE 
~\cite{itse} also achieves an accurate, lightweight text-to-speech alignment module implemented without expertise such as pronunciation lexica. 

These works, however, conduct force-alignment outside of ASR models to get timestamps. For an ASR model which is required to obtain timestamp prediction ability, such models introduce additional computation overhead, which might be unacceptable for ASR systems in commercial usage. The excepted timestamp prediction models inside ASR system is supposed to introduce less computation overhead and predict accuracy timestamps naturally.

\section{Preliminaries}

\subsection{Continuous Integrate-and-Fire}

\begin{figure}[htbp]
    \centering
    \includegraphics[width=0.6\linewidth]{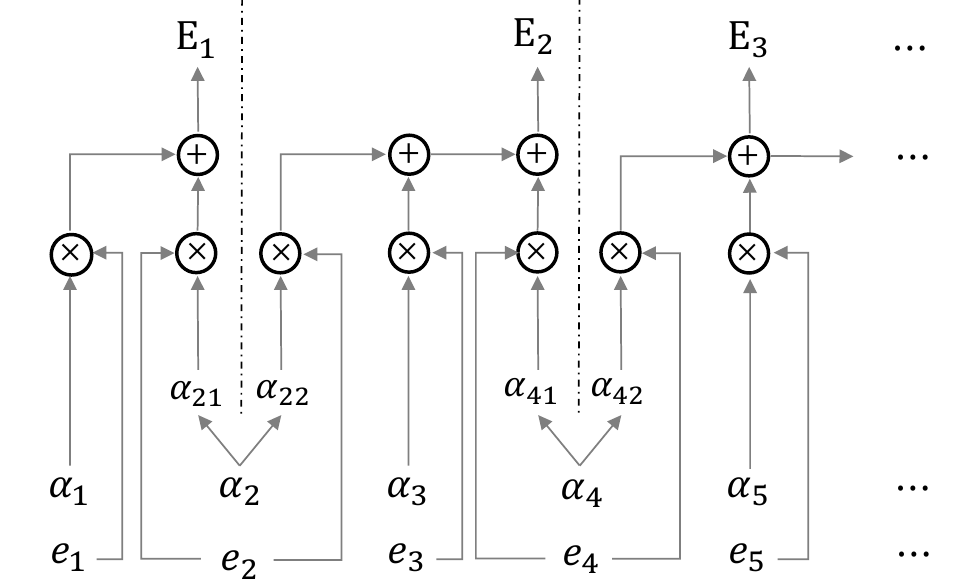}
    \caption{Illustration of integrate-and-fire on encoder output $\mathbf{e}_{1:T}$ with predicted weights $\alpha=(0.3,0.9,0.4,0.4,0.3)$. The integrated acoustic embedding $\mathbf{E}_{1}=0.3\times\mathbf{e}_{1}+0.7\times\mathbf{e}_{2}$, $\mathbf{E}_{2}=0.2\times\mathbf{e}_{2}+0.4\times\mathbf{e}_{3}+0.4\times\mathbf{e}_{4}$. The sum of weights $\alpha$ is $L^{'}$ - the length of prediction sequence.}\label{cif}
\end{figure}

Continuous integrate-and-fire~(CIF) is a soft and monotonic alignment mechanism for E2E ASR different from time-synchronized models and label-synchronized models. CIF performs integrate process over the output of encoder $\mathbf{e}_{1:T}$ and predicts frame-level weights $\mathbf{\alpha}_{1:T}$, once the accumulated weights exceed the fire threshold, these frames will sum up to acoustic embedding $\mathbf{E}_{1:L^{'}}$ which is synchronized with output tokens. Such process is illustrated in Fig~\ref{cif}. In ASR models, the modeling character of CIF is of great value. The frame-level weights actually conduct  alignment between acoustic representation and output tokens, and the fire location indicates the token boundary. Such capacity delights us to explore the feasibility of using CIF to predict accurate timestamps of decoded tokens naturally in the process of recognition.

\subsection{Paraformer}

\begin{figure}[htbp]
    \centering
    \includegraphics[width=0.6\linewidth]{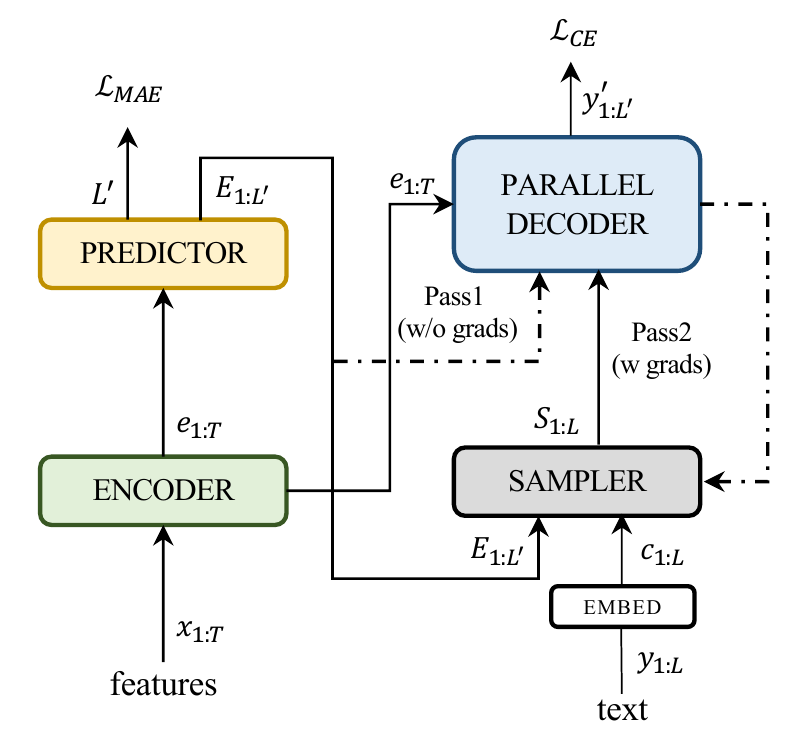}
    \caption{Illustration of Paraformer structure.}\label{para}
\end{figure}

We adopt Paraformer - a novel NAR ASR model which achieves non-autoregressive decoding capacity by utilizing CIF and two-pass training strategy inside an AED backbone. Avoiding the massive computation overhead introduced
by autoregressive decoding and beam-search, Paraformer gains more than 10x speedup with even lower error rate comparing to Conformer baseline. The overall framework of Paraformer is illustrated in Fig~\ref{para}.

Paraformer contains three modules, namely encoder, predictor and parallel decoder~(sampler interpolates acoustic embeding and char embedding without parameters). Encoder is same as AR encoders of Conformer which contains self-attention, convolution and feed-forward networks~(FFN) layers to generate acoustic representation $\mathbf{e}_{1:T}$ from down-sampled Fbank features. Predictor uses CIF to predict the number of output tokens $L^{'}$ and generate acoustic embedding $\mathbf{E}_{1:L^{'}}$. Parallel decoder and sampler conduct two-pass training with the vectors above: $\mathbf{E}_{1:L^{'}}$ is directly sent to decoder to calculate cross-attention with $\mathbf{e}_{1:T}$, and decode all tokens $y^{'}_{1:L^{'}}$ in NAR form at once. Then sampler interpolates $\mathbf{E}_{1:L^{'}}$ and char embedding $\mathbf{c}_{1:L}$ according to the edit distance between hypothesis $y^{'}_{1:L^{'}}$ and $y_{1:L}$. The interpolated vector (named semantic embedding, noted by $S_{1:L^{'}}$) is sent to parallel decoder again to conduct the same calculation, which makes up the second pass. Note that the forward process of the first pass is gradient-free, cross-entropy~(CE) loss is calculated between output of the second pass and ground truth. In the training process, as the accuracy of first pass decoding raises, $\mathbf{E}_{1:L^{'}}$ makes up increasing proportion of $S_{1:L^{'}}$. In the inference stage, Paraformer use the first pass to decode.

\begin{figure}[htbp]
    
    \includegraphics[width=\linewidth]{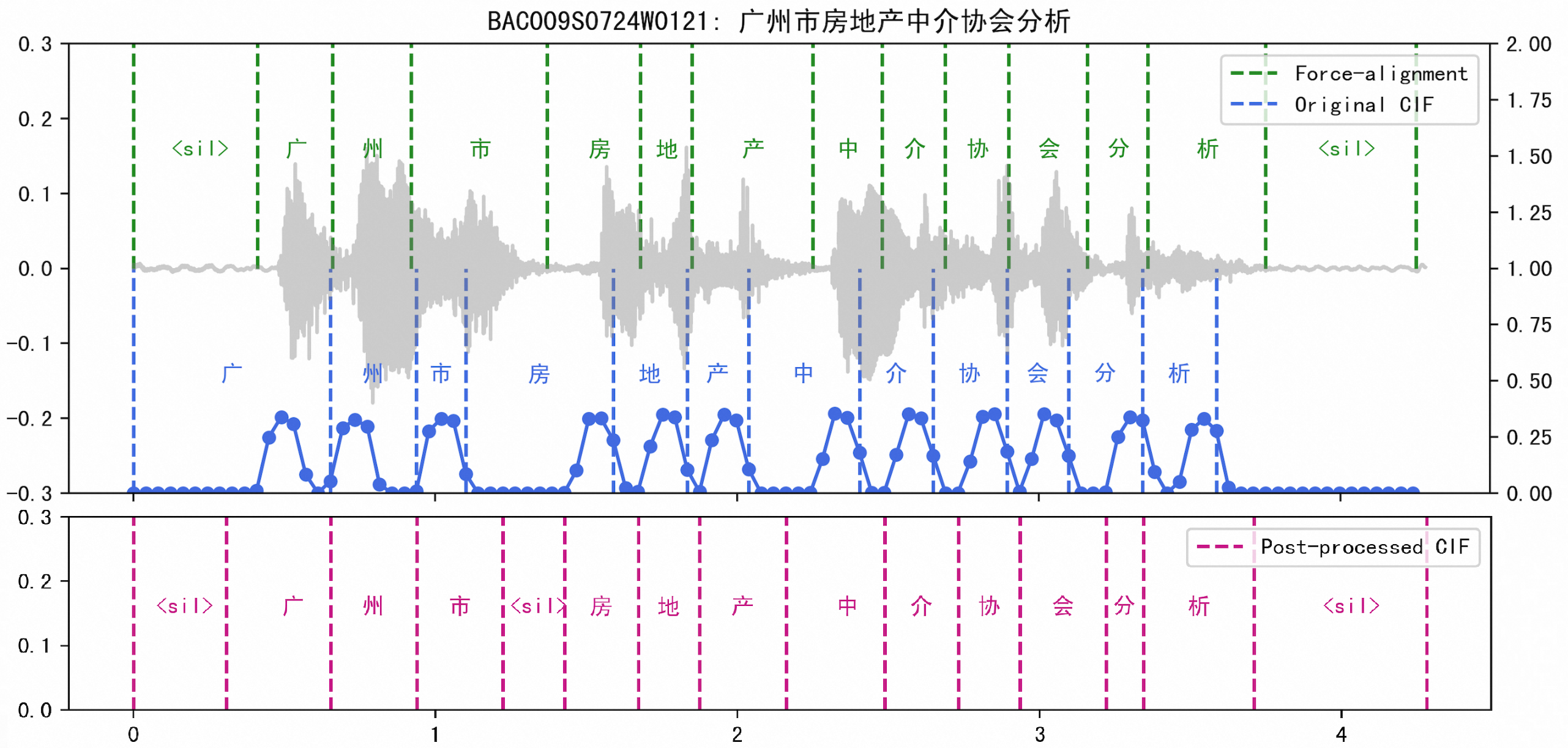}
    \caption{CIF fire places and weights $\alpha$ of demo utterance from Aishell-1. Axis of $\alpha$ is in the right. The subfigure above shows the timestamps of origin CIF comparing to FA system, the CIF fire place roughly indicate the corresponding token place but it is always inaccurate, especially the end of integrate is usually later. The subfigure at bottom shows the effect of weights post-processing strategies introduced in section 4.2.}\label{tsshow}
\end{figure}

CIF is of vital importance in Paraformer as it predicts the number of output tokens and generates monotonic token boundaries implicitly. Fig~\ref{tsshow} shows the origin fire place comparing to the timestamp generated by kaldi force-alignment system. Weights $\alpha$ shows that the pattern of CIF is regard less of the real length of tokens, for an ASR model with 4-times down-sampling layer~(60 ms each time step), the integrate process for each token finishes in around 4 frames, which leads to a large offset for end point timestamp of character with long duration.

\section{Methods}

Considering of the pattern of CIF weights and the characteristic of Paraformer, we optimize the timestamp prediction strategy from two aspects. In the training process, we propose to scale the CIF weights after sigmoid function in order to alleviate the sharpness of weights and also cut off the gradient towards encoder of irrelevant position. Then we adopt several post-processing strategies for weights to achieve more precise timestamp prediction.

\subsection{Scaled-CIF Training}

Original CIF calculates $\alpha$ with sigmoid activation function after feed-forward network, sharp spikes and glitches can be observed from the curve in Fig ~\ref{tsshow}. We propose to scale and smooth the CIF weights with the operation below:

\begin{equation}
    \alpha^{'}=\gamma \cdot ReLU\big((sigmoid(x)-\beta\big)
\end{equation}

First, ReLU function and $\beta$ smooth the glitches of $\alpha$, cutting off the gradient towards encoder of irrelevant position, which is supposed to be beneficial for ASR task. Besides, scaling the output of ReLU with $\gamma$ relieves the spikes of the curve, trying to achieve level and smooth weights.

\subsection{Weights Post-Processing}

Another optimization comes about from the aspect of post-processing. According to the distribution of alphas observed in Fig~\ref{tsshow}, most of the weights begins to accumulate from the exact frame but the accumulation ends in fixed steps~(4 frames in the figure), which delights us to conduct the following 4 processing strategies:

\textbf{Begin/end Silence} As the begging place of CIF is always precise, the beginning frames with weights under a threshold $\theta_s$ is considered as silence. So as to the frames at the end but we set an interval of 3 frames.

\textbf{Fire Delay} Original CIF timestamps are always unreliable for the end point prediction for long-lasting tokens. We propose to conduct fire delay operation when frames with low weights are observed, such frames with be grouped to the previous token except the last frame.

\textbf{Silence Insertion} When the low-weight frames last longer than $L_s$, we insert a silence token in between.

\textbf{Weight Averaging\footnote{In this paper, we conduct experiments with 4-times down-sampling encoder, this methods is thus not validated.}} For ASR models with higher down-sampling rate, the process of integrate finishes even faster. When using 6-times low frame rate features, the CIF tends to output weights around 1.0 or 0.0, which makes CIF weights no longer stand for the accumulation procedure. Then we proposed to weaken the weight spikes.

$\theta_s$ and $L_s$ are model related hyper-parameters, for Paraformer model with original CIF predictor and 4-times down-sampling encoder embedder, we set $\theta_s=0.05$ and $L_s=3$, the timestamp thus generated of the demo utterance above is shown in Fig~\ref{tsshow}. Subjectively, the proposed post-processing strategies optimize the quality of CIF timestamps in a simple but efficient way.

\subsection{Evaluation Metrics}

In addition to the visual analysis of timestamps, we propose to use accumulated averaging shift~(AAS) and diarization error rate~(DER) as the evaluation metrics of timestamp accuracy.

\noindent \textbf{AAS} The first metrics measures the averaging time shift of each token. The time shift of begging timestamp and end timestamp are summed up and averaged over the entire testset. Formally, we calculate the metrics between timestamp $i$ and timestamp $j$ as

\begin{equation}
AAS=\frac{\sum_{k=1}^K\left|start_{k, i}-start_{k, j}\right|+\left|e n d_{k, i}-e n d_{k, j}\right|}{2K},
\end{equation}
where $K$ is for the number of aligned token pairs\footnote{Considering the transcriptions of two timestamps might differ, only paired tokens according to edit distance are included in the calculation.}.

\noindent \textbf{DER} Speaker diarization systems are evaluated by DER, which calculates the proportion of frames  which are classified correctly. DER is the sum of
three different error types: False alarm of speech, missed detection of speech and confusion between speaker labels. Treating tokens as speakers, we introduce this metrics to measure the quality of timestamp.

\begin{equation}
DER=\frac{False~Alarm+Missed+Speaker~Confusion}{Total~Duration~of~Time} .
\end{equation}

\section{Experiments and Results}

\subsection{Datasets}

Two datasets are used in our experiments. First, Aishell-1 is used for training the Paraformer model and visualizing timestamps. Aishell-1 contains 178 hours speech data with transcription, which is a widely used open-source Mandarin ASR corpus.
Besides, we use a TTS dataset called M7 for evaluation, which contains 5550 Mandarin utterances. Except transcriptions, M7 also contains manually marked timestamps in token level, which is regarded as reference in the calculation of AAS and DER.

\subsection{Experiment Setup}

The Paraformer model is trained from scratch using ESPnet toolkit with the following setups. The model contains 12-layer Conformer encoder~(implemented as with kernel size 15) and 6-layer Transformer decoder with attention dimension $d_{attn}=256$ and feed-forward network dimension $d_{linear}=2048$. The input layer of encoder conducts 4-times down-sampling for Fbank features, one step of encoder output thus stands for 40ms. The model is jointly trained with CTC loss~($\alpha=0.3$) and we set dropout rate $r_{dropout}=0.1$ for entire model. 3-times speed perturbation is adopted for Aishell-1 data and we apply spec-augment with 2 frequence masks range in $[0, 30]$ and 2 time masks range in $[0, 40]$ for each utterance. We use dynamic batch size~($numel=2000k$) and $n_{epoch}=50$. No language model is used in the inference stage. For scaled-CIF, we use $\gamma=0.8$ and $\beta=0.05$\footnote{For ASR models of different frame rate and encoder down-sampling rate, we fine it better to adjust scaling coefficients $\gamma$ and $\beta$, so as to the post-processing hyper-parameters $\theta_s$ and $L_s$ to achieve better performance.}.

We prepare the force-alignment system with Kaldi toolkit as baseline timestamp, the training setup and model configuration is exactly same as open-source recipe Aishell-1, from flat-start to HMM-GMM. Timestamps of phonemes are extracted from lattice using model \textit{tri5a}, and then converted to characters timestamps.

\subsection{Quality of Timestamp}

In this section we evaluate the quality of the timestamps and analyse the effect of the proposed methods. Table~\ref{res} shows the AAS and DER metircs of the timestamps from different models over testset M7. First we test the force-alignment system with ground truth transcription and Paraformer recognition results. It turns out that using these two kinds of transcription lead to slight difference of DER and almost the same AAS, which is because the majority error of Paraformer's recognition results is substitution error, reference and hypothesis have nearly the same expanded phoneme sequences, thus AAS of FA-GT and FA-HYP differs little.

\begin{table}[htbp]
\centering
\caption{Accuracy of timestamp from force-alignment system and Paraformer CIF.}\label{res}
\begin{tabular}{lllllll}
\hline
                                         &  & Exp    & Sys                                        &  & AAS~(sec)   & DER~(\%)   \\ \cline{1-1} \cline{3-4} \cline{6-7} 
\multirow{2}{*}{force-alignment systems} &  & FA-GT  & FA with groundtruth                        &  & 0.080 & 6.34  \\
                                         &  & FA-HYP & FA with decoded trans                      &  & 0.081 & 7.50  \\ \cline{1-1} \cline{3-4} \cline{6-7} 
\multirow{4}{*}{CIF timestamps}          &  & CIF-0  & origin CIF timestamp                       &  & 0.213 & 45.39 \\
                                         &  & CIF-1  & +begin/end silence                         &  & 0.161 & -     \\
                                         &  & CIF-2  & ~~+fire-delay                    &  & 0.124 & -     \\
                                         &  & CIF-3  & ~~~~+silence insertion &  & 0.112 & 17.09 \\ \cline{1-1} \cline{3-4} \cline{6-7} 
\multirow{4}{*}{scaled-CIF timestamps}   &  & SCIF-0 & scaled-CIF timestamp                       &  & 0.143 & 29.75 \\
                                         &  & SCIF-1 & +begin/end silence                         &  & 0.098 & -     \\
                                         &  & SCIF-2 & ~~+fire-delay                    &  & 0.080 & -     \\
                                         &  & SCIF-3 & ~~~~+silence insertion &  & 0.071 & 8.11  \\ \hline
\end{tabular}
\end{table}

Comparing CIF-0 and FA-HYP, it is obvious that original CIF weights as timestamps is of unacceptable quality. With the addition of post-processing strategies, the accuracy of CIF timestamps has been improved step by step, CIF-3 achieves 47.4\% AAS reduction. Among the three post-processing methods, fire delay is the most effective and also the most tricky one. Scaled-CIF brings further help to timestamps prediction, 32.9\% AAS reduction and 34.5\% DER reduction are observed without any post-processing strategies. In SCIF-3, CIF timestamps outperforms force-alignment systems in AAS~(11.3\% relatively), but DER is still higher.

\begin{figure}[htbp]
    \centering
    \includegraphics[width=\linewidth]{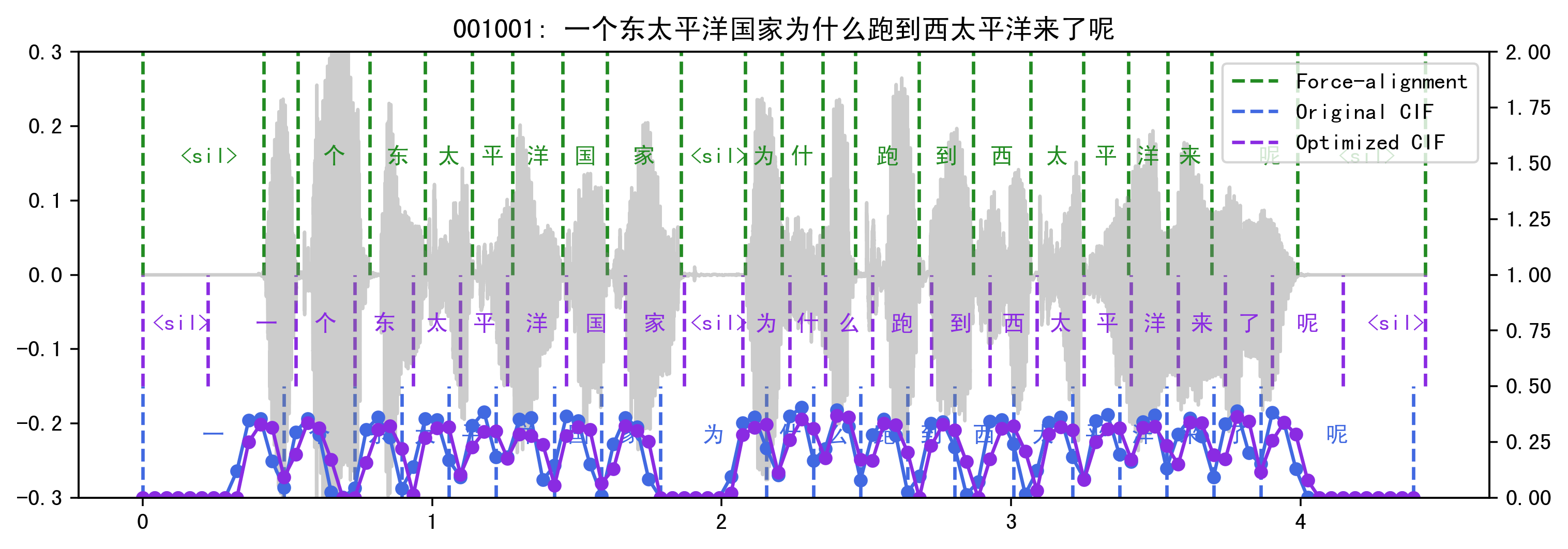}
    \caption{Demo for utterance 001001 in M7. It can be observed that the peak in the $\alpha$ is weakened and the optimized timestamps are more accurate.}\label{001001}
\end{figure}

Fig~\ref{001001} shows comparison of manually marked timestamps, original CIF timestamps and optimized timestamps. Comparing the blue curve and the purple curve, it can be observed that the peak of the curve is weakened to some extent, and the glitches in the curve disappear~(not obvious in the figure). Considering the start and ending time of each token, the effect of fire delay is obvious, especially when the token is followed by low weight frames~(like character \begin{CJK*}{UTF8}{gbsn}'家'\end{CJK*} in the demo).

\subsection{ASR Results}

Results in section 5.3 shows that scale-CIF significantly improves timestamp prediction accuracy, and we find it is also benificial for ASR.
Comparing to vanilla Conformer AR model, Paraformer achieves better recognition accuracy and even lower real time factor~(RTF)~\cite{paraformer}. On Aishell-1 task, Paraformer gets 4.6\%, 5.2\% CER over dev and test set and the RTF is 0.0065~(RTF of Conformer is 0.1800 under the same test environment). In our experiments, Paraformer with scaled-CIF outperforms the baseline, CER on Aishell-1 dev and test set are 4.5\% and 5.2\% respectively. On testset M7, Paraformer gets 11.70\% CER with origin CIF and 11.26\% with scaled CIF~(3.76\% relative CER reduction). Such results prove that smoothing the weights of CIF and cutting off the gradient towards encoder of irrelevant frames are beneficial for recognition task of Paraformer.

\section{Conclusion}

As the inherent deficiency of predicting timestamps with AED based models, we propose to predict timestamps according to CIF weights while recognition with Paraformer. In this paper, we first explore the characteristic of CIF in Paraformer for Mandarin. It turns out that the integrate of CIF weights tends to start from the right place but the process ends in fixed number of steps, regardless of tokens' real length. Besides, sharp peaks and glitches are observed in CIF weights. Such behavior delights us to improve the accuracy of CIF timestamp with scaled-CIF and post-processing strategies. We compare the CIF timestamp with force-alignment results of conventional HMM-GMM systems, and evaluate the quality of timestamps with AAS and DER metrics. The results in Table~\ref{res} show that with the help of the proposed methods, CIF timestamps achieve comparable performance as the FA baseline, 12.3\% relative AAS reduction is observed while DER is a little worse. Comparing CIF timestamps before and after optimization, 66.7\% AAS reduction and 82.1\% DER reduction is achieved. In summarize, the proposed scaled-CIF and post-processing strategies improve the accuracy of timestamp and out system outperforms the conventional hybrid model trained with the same data, and such system reduces computation overhead which might be of value in commercial usage.

In the future, we will explore the CIF timestamps of different languages and different frame-rate, and modify the silence insertion strategy with dynamic silence length threshold according to the phoneme.

%
%
%
\bibliographystyle{splncs04}
%

\end{document}